\documentclass[]{article}
\usepackage[a4paper, total={6in,8in}]{geometry}
\usepackage[affil-it]{authblk}
\usepackage{amsmath}
\usepackage{mathtools}
\usepackage{graphicx}
\usepackage{subcaption}
\usepackage{setspace}
\usepackage{natbib}

\begin{document}

\title{Generalized Entropy Approach for Conserved Systems with Finite versus Infinite Entities: Insights into Non-Gaussian and Non-Chi-Square Distributions using Havrda-Charv\'at-Tsallis Entropy versus Gaussian Distributions via Boltzmann-Shannon Entropy}
\author{Jae Wan Shim}
\affil{Extreme Materials Research Center, Korea Institute of Science and Technology, and\\Division of AI-Robotics, KIST Campus, University of Science and Technology, \\5 Hwarang-ro 14-gil, Seongbuk, Seoul 02792, Republic of Korea}
\date{}
\maketitle

\doublespacing
\large

\begin{abstract}
We demonstrate that the most probable state of a conserved system with a limited number of entities or molecules is the state where non-Gaussian and non-chi-square distributions govern. We have conducted a thought experiment using a specific setup. We have verified the mathematical derivation of the most probable state accurately predicts the results obtained by computer simulations. The derived distributions approach the Gaussian and the chi-square distributions as the number of entities approaches infinity. The derived distributions of the most probable state will have an important role in the fields of medical research where the number of entities in the system of interest is limited. Especially, the non-chi-square distribution can be interpreted by an asset distribution achieved after a repetitive game where an arbitrary portion of one's assets is transferred to another arbitrary entity among a number of entities with equal abilities.
\end{abstract}


\section{Introduction}
\label{sec:intro}
Utilizing the concept of infinity, the central limit theorem allows us to use Gaussian and chi-square distributions instead of exact distributions under various circumstances. In our study, however, we examine exact distributions of non-Gaussian and non-chi-square that can be derived through rigorous mathematical methods.

The non-Gaussian and non-chi-square distributions are widespread across various fields including economics, plasma physics, atmospheric sciences, and semiconductor engineering. In economics, non-Gaussian distributions have been observed in the variation of speculative prices by \cite{Mandelbrot} and the empirical properties of asset returns by \cite{Cont}. Similarly, extensive scientific studies \citep{Chamberlain, Hubert, Lockwood, Pilipp, Kaniadakis, Hori, Maksimovic, Benedetto,  Gehring,  Rubab1, Rubab,  Zaheer, Zouganelis} and statistical modeling of extreme values of \cite{Coles} have demonstrated the existence of non-Gaussian distributions in other fields.

The investigation of non-Gaussian and non-chi-square distributions and their underlying principles is of great importance, as it enables us to understand complex systems and their behaviors.
In this study, we present a rigorous derivation of velocity and energy distribution functions for a system of limited number of entities where the definition of entities includes individual participants. By deriving these distribution functions, we provide insights into the statistical behavior and its relevant entropies of such systems, with potential applications in fields ranging from physics to economics.

\section{Derivation of energy or asset distribution}
\label{sec:derivation}

To derive the exact distribution for a system with a limited number of entities, we start by using an equation inspired by the approach of \cite{cercignani1969} for deriving the Maxwellian velocity distribution. While Cercignani used velocities as independent variables, we employ energies as independent variables in our equation to emphasize that the distribution derived in this paper is applicable to systems that conserve quantities with respect to both a square sum and a direct sum of entity values, such as an economic system with a fixed global asset analogous to a fixed global energy.

Let us define a space $\mathbf S$ and a region $\mathbf R \in \mathbf S$, which represents a conserved system with respect to the number of entities $N$ and a given energy that is bounded. Using this definition, we can write an energy distribution function $f_\epsilon$ as

\begin{equation}
\label{eq:DiracEnergy}
f_\epsilon(\epsilon_1, \cdots, \epsilon_N)=c_\epsilon \, \delta \left( \sum_{i=1}^N \epsilon_i  -E\right)
\end{equation}
where $\epsilon_i$ is the energy of the $i$th molecule, $c_\epsilon$ is a normalizing constant, $\delta(\cdot)$ is the Dirac delta function, and $E$ is a given energy of the system in $\mathbf R$.

To facilitate the integral calculations of the Dirac delta function, which will be done later, let us define variables $q_i$ by
$$ q_i=\pm \sqrt \epsilon_i $$
which is purely a mathematical change of variables at this stage and alternatively we can define $q_i=\sqrt \epsilon_i$. The choice does not affect the final result of one-entity energy distribution. Let us define a distribution function with respect to $q_i$ as
\begin{equation}
\label{eq:DiracQ}
f_q(q_1, \cdots, q_N)=c_q \, \delta \left( \sum_{i=1}^N q_i^2  -E\right)
\end{equation}
where $c_q$ is another normalizing constant. Then, it is convenient to deal with the summation in the delta function by adopting the polar coordinate system as the followings,
$$q_i = r \cos \phi_i \prod_{j=1}^{i-1} \sin \phi_j  $$
for $i=1, \cdots, N-1$ and
$$q_N = r \prod_{j=1}^{N-1} \sin \phi_j $$
where the ranges of the variables are defined by $ 0 \leq r \leq \infty$ and $0 \leq \phi_j \leq \pi$ for $j=1, \cdots, N-2$ and $0 \leq \phi_{N-1} \leq 2\pi$. Now, we have a compact expression 
$$\sum_{i=1}^N {q}_i^2=r^2$$ 
and the infinitesimal volume $dq_1 \cdots dq_N$ can be written as 
$$dq_1 \cdots dq_N=r^{N-1} dr \, d\mathbf A_{N}$$
where $\mathbf A_{N}$ is the surface area of the $N$-dimensional unit ball and its value is given by 
$$\mathbf A_{N}=\frac {2 \pi^{N/2}}  {\Gamma \left(\frac N 2\right)}$$
where $\Gamma(\cdot)$ is the gamma function defined by using the double factorial as
$$ \Gamma \left(\frac N 2\right) = 2^{\left(- \frac{N-1}{2} \right)} (N-2)!! \sqrt \pi.$$

Then, the normalizing constant $c_q$ is obtained by using the following equation,
\begin{equation}
\label{eq:NormalizingQ}
c_q \int \delta \left(r^2 - E \right) r^{N-1} dr \int d {\mathbf A}_N=\frac {c_q} {2} {\mathbf A}_N  E^{\frac {N-2}{2}}=1,
\end{equation}
which is
$$c_q=\frac 2 {{\mathbf A}_N E^{\frac {N-2}{2}}.}$$ 
When we integrate Eq.~(\ref{eq:DiracQ}) with respect to $q_i$ for $i=1, \cdots, N-1$ with excluding $q_N$, the result of the integration becomes as the following;
\begin{align}
\label{eq:OneMoleculeQ}
F_q(q_N) & =\int \cdots \int f_q({q}_1, \cdots, {q}_N) dq_1 \cdots dq_{N-1} \nonumber \\ 
            & =c_q \, {\mathbf A}_{N-1} \int \delta \left(r^2 -( E-q_N^2)\right)r^{N-2}dr \nonumber \\
            & = 
		\begin{cases} 
   			{{\mathbf A}_{N-1}} \left({\mathbf A}_{N} E ^ {\frac{N-2}{2}}\right)^{-1}(E-q_N^2)^{\frac{N-3}{2}} & \text{if } q_N^2 \leq E \\
   			0       & \text{if } q_N^2 >  E.
  		\end{cases}
\end{align}

Suppose the energy has no preference for any particular entity. In that case, the distribution function must exhibit symmetry with respect to changes in the index of entities, allowing us to eliminate the subscript $N$ from $q_N$ for the sake of simplicity. By substituting $\mathbf A_{N-1}$ and $\mathbf A_{N}$ with the expressions of the Gamma function, we have
\begin{equation}
\label{eq:OneMoleculeQSimple}
F_q(q)=\frac{\Gamma \left(\frac N 2 \right)} {\Gamma \left(\frac {N-1} 2 \right) \sqrt {\pi E}}\left(1-\frac{q^2}{E} \right)^{\frac {N-3}{2}}
\end{equation}
for $q^2 \leq E$ and otherwise $F_q(q)=0$.

Let us define the one-entity energy distribution $F_\epsilon(\epsilon_N)$ by using Eq.~(\ref{eq:DiracEnergy}) as
$$
F_\epsilon(\epsilon_N) 
	=\int \cdots \int f_\epsilon({\epsilon}_1, \cdots, {\epsilon}_N) d\epsilon_1 \cdots d\epsilon_{N-1}
$$
and with considering the following identity
\begin{equation}
\int_0^\infty F_\epsilon(\epsilon) d\epsilon = 2 \int_0^\infty F_q(q) dq 
\end{equation}
with $dq=(2\sqrt \epsilon)^{-1} d\epsilon$, the energy distribution $F_\epsilon(\epsilon_N)$ or simply $F_\epsilon(\epsilon)$ is, then, obtained by  
\begin{align}
\label{eq:OneMoleculeEnergy}
F_\epsilon(\epsilon) 
	&=\frac{1}{\sqrt {\epsilon_N}}F_q(q)\nonumber\\
	&=\frac{\Gamma \left(\frac N 2 \right)} {\Gamma \left(\frac {N-1} 2 \right) \sqrt {\pi \epsilon E}}\left(1-\frac{\epsilon}{E} \right)^{\frac {N-3}{2}}
\end{align}
for $\epsilon \leq E$ and otherwise $F_\epsilon(\epsilon)=0$.

\begin{figure}
\centering
\includegraphics[scale=0.9]{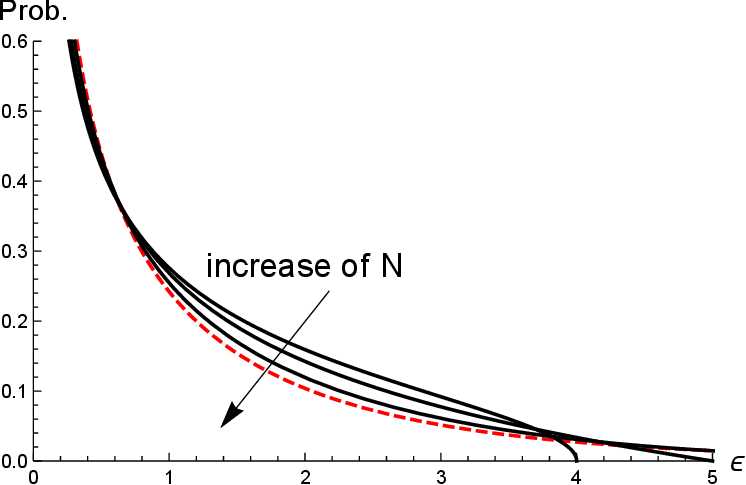}
\caption{Graphs of $F_{\bar \epsilon}(\bar \epsilon)$ for $N \in \{4,5,10\}$ and $\chi^2_{\bar \epsilon}(\bar \epsilon)$ defined by Eqs.~(\ref{eq:OneMoleculeEnergyAdimensional}) and (\ref{eq:chiSquareAdimensional}). The solid lines (black) are the graphs of $F_{\bar \epsilon}(\bar \epsilon)$ when $N=4$, $5$, and $10$; and the dashed line (red) is the graph of $\chi^2_{\bar \epsilon}(\bar \epsilon)$. Note that $F_{\bar \epsilon}(\bar \epsilon)$ approaches $\chi^2_{\bar \epsilon}(\bar \epsilon)$ as $N$ increases.}
\label{fig:fig0}
\end{figure}

As $N$ approaches infinity, Eq.~(\ref{eq:OneMoleculeEnergy}) approaches the Maxwellian energy distribution, which is a chi-square distribution with one degree of freedom as
\begin{equation}
\label{eq:chiSquare}
\chi^2_\epsilon(\epsilon) = \frac{1} {\sqrt {2 \pi \epsilon \bar E}}\exp \left (-\frac{\epsilon}{2 \bar E} \right)
\end{equation}
where the total energy $E$ is assumed to be linearly dependent on $N$ as $E= N \bar E$. One can use $\bar E= (k T)/2$ from the following relation of the kinetic theory of gases that the sum of the kinetic energies of molecules in one-dimensional space can be expressed by
$$E=\frac m 2 \sum_i {v}_i^2=\frac 1 2 N k T$$
where $v_i$ is the velocity of the $i$th molecule, $k$ is the Boltzmann constant, and $T$ is a temperature. Then, Eq.~(\ref{eq:chiSquare}) becomes
\begin{equation}
\label{eq:MaxwellianEnergy}
M_\epsilon(\epsilon) = \frac{1} {\sqrt {\pi \epsilon k T}}\exp \left (-\frac{\epsilon}{k T} \right).
\end{equation}
By defining $\epsilon = \bar E \bar \epsilon$, we respectively obtain dimensionless versions of Eqs.~(\ref{eq:OneMoleculeEnergy}) and (\ref{eq:chiSquare}) as
\begin{equation}
\label{eq:OneMoleculeEnergyAdimensional}
{F_{\bar \epsilon}(\bar \epsilon)} 
	=\frac{\Gamma \left(\frac N 2 \right)} {\Gamma \left(\frac {N-1} 2 \right) \sqrt {\pi \bar \epsilon N}}\left(1-\frac{\bar \epsilon}{N} \right)^{\frac {N-3}{2}}\end{equation}
and
\begin{equation}
\label{eq:chiSquareAdimensional}
{\chi^2_{\bar \epsilon}(\bar \epsilon)} = \frac{1} {\sqrt {2 \pi \bar \epsilon}}\exp \left (-\frac{\bar \epsilon}{2} \right).
\end{equation}
The graphs of Eqs.~(\ref{eq:OneMoleculeEnergyAdimensional}) and (\ref{eq:chiSquareAdimensional}) are shown in Fig.~(\ref{fig:fig0}). The solid (black) lines correspond to the graphs of Eq.~(\ref{eq:OneMoleculeEnergyAdimensional}) for $N=4$, $5$, and $10$; and the dashed (red) line corresponds to the graph of Eq.~(\ref{eq:chiSquareAdimensional}). Note that the mean of $\bar \epsilon$ with respect to ${F_{\bar \epsilon}(\bar \epsilon)}$ is $1$ and the variance is $2(N-1)/(N+2)$.
 
\section{Derivation of velocity distribution}
\label{sec:derivation2}

It is convenient to start from Eq.~(\ref{eq:OneMoleculeQSimple}) to show a velocity distribution. By letting
$$ q_i=\sqrt{\frac m 2} v_i\,, $$
we have
\begin{equation}
\label{eq:OneMoleculeVelocity}
F_v(v) =\frac{\Gamma \left(\frac N 2 \right) \sqrt m} {\Gamma \left(\frac {N-1} 2 \right) \sqrt {\pi N k T}}\left(1-\frac{m v^2}{N k T} \right)^{\frac {N-3}{2}}
\end{equation} 
and it is straightforward to show that Eq.~(\ref{eq:OneMoleculeVelocity}) approaches the Gaussian distribution, \textit{i.e.} the Maxwellian velocity distribution, as $N$ increases. By defining
$$ g(v)=\left(1-\frac{mv^2}{NkT} \right)^{\frac {N-3}{2}},$$
we have
$$ \lim_{N \to \infty} \ln g(v) = - \frac {mv^2}{2kT}.$$
And also we have
$$  \lim_{N \to \infty} \frac{\Gamma \left(\frac N 2 \right) \sqrt m}{\Gamma \left(\frac {N-1} 2 \right) \sqrt {\pi N k T}}=\frac {\sqrt m}{\sqrt {2\pi k T}}$$
so that we can obtain the molecular velocity distribution when $N$ approaches infinity as
$$ M(v)= \lim_{N \to \infty} F_v(v)=\sqrt{\frac { m}{ {2\pi k T}}} \exp \left(- \frac {mv^2}{2kT} \right)$$
which is the Maxwellian velocity distribution in one-dimensional space.

\begin{figure}
\centering
\includegraphics[scale=0.9]{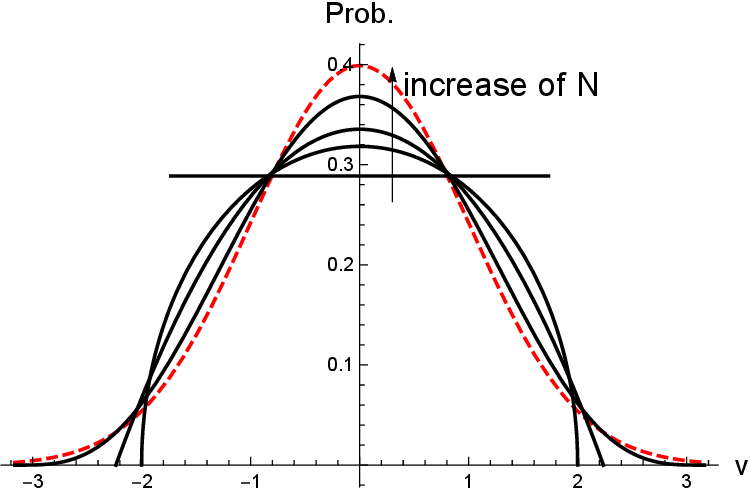}
\caption{Graphs of $F_{\bar v}(\bar v)$ for $N \in \{3,4,5,10\}$ and $\bar M(\bar v)$ defined by Eqs.~(\ref{eq:OneMoleculeVelocityAdimensional}) and (\ref{eq:MaxwellianVelocityAdimensional}). The straight line (black) is the graph of $F_{\bar v}(\bar v)$ when $N=3$ and the curved lines (black) are the graphs when $N \in \{4,5,10\}$. The dashed line (red) is the graph of $\bar M(\bar v)$. Note that $F_{\bar v}(\bar v)$ approaches $\bar M(\bar v)$ as $N$ increases.}
\label{fig:fig1}
\end{figure}

Let us define a dimensionless velocity $\bar v = v \sqrt{{m}/{(kT)}}$. Then, we have
\begin{equation}
\label{eq:OneMoleculeVelocityAdimensional}
{F_{\bar v}(\bar v)}=\frac{\Gamma \left(\frac N 2 \right) }{\Gamma \left(\frac {N-1} 2 \right) \sqrt {\pi N}}\left(1-\frac {\bar v^2}{N} \right)^{\frac{N-3}{2}}
\end{equation}
and
\begin{equation}
\label{eq:MaxwellianVelocityAdimensional}
\bar M(\bar v) = {\frac { 1}{ \sqrt{2\pi}}} \exp \left(- \frac {\bar v^2}{2} \right).
\end{equation}
In Fig.~(\ref{fig:fig1}), we present the graphs of $F_{\bar v}(\bar v)$ for $N \in \{3,4,5,10\}$ and $\bar M(\bar v)$. As $N$ increases, $F_{\bar v}(\bar v)$ approaches the Gaussian distribution $\bar M(\bar v)$. 

For the case of the Gaussian distribution, the expectation values of the square of velocity and the speed are respectively obtained by
$$ {\mathbf E}_M(v^2) = \int v^2 M(v) dv = \frac{kT}{m}$$
and
$$ {\mathbf E}_M(|v|) = \int |v| M(v) dv =\sqrt {\frac{2kT}{\pi m}}.$$ 

Meanwhile, for the case of the non-Gaussian distribution $F_v(v)$, the expectation values are respectively obtained by
$$ {\mathbf E}(v^2) = \int v^2 F_v(v) dv = {\mathbf E}_M(v^2)$$
and
$$ {\mathbf E}(|v|) = \int |v| F_v(v) dv =\sqrt {\frac{N}{2}}{\frac{\Gamma(N/2)}{\Gamma \left((N+1)/2\right)}}{\mathbf E}_M(|v|).$$

\begin{figure}
\centering
\includegraphics[scale=0.9]{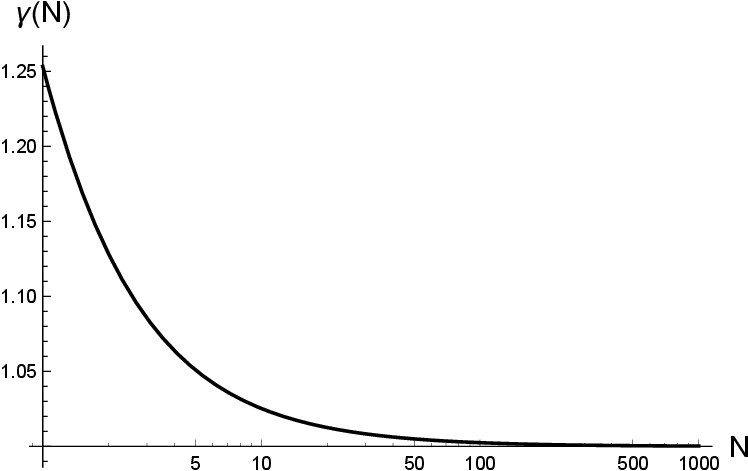}
\caption{Log-linear plot of $\gamma(N)$ with respect to $N$ in Eq.~(\ref{eq:Gamma}). The ratio $\gamma(N)$ approaches $1$ as $N$ increases.}
\label{fig:fig2}
\end{figure}

We define
\begin{equation}
\label{eq:Gamma}
\gamma(N) = \frac {{\mathbf E}(|v|)}{{\mathbf E}_M(|v|)}= \sqrt {\frac{N}{2}}{\frac{\Gamma(N/2)}{\Gamma \left((N+1)/2\right)}}
\end{equation}
and plot $\gamma(N)$ with respect to $N$ in Fig.~(\ref{fig:fig2}). The $\gamma(N)$ that is the ratio between ${\mathbf E}(|v|)$ and ${\mathbf E}_M(|v|)$ rapidly approaches $1$ as $N$ increases.

\section{Verification}
\label{sec:verification}

\begin{figure}
\centering
\includegraphics[scale=0.6]{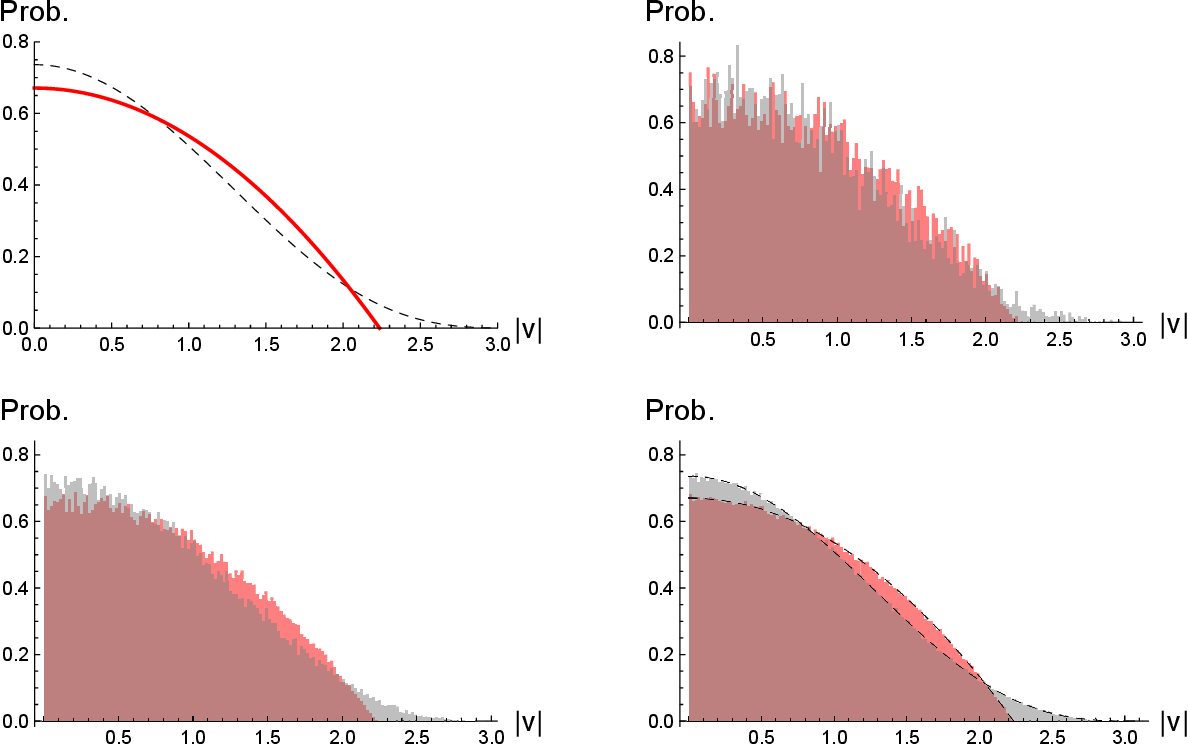}
\caption{Speed distributions obtained through theoretical methods and computer simulations are presented in this figure. The upper-left panel shows a reference distribution calculated using Eq.~(\ref{eq:OneMoleculeVelocityAdimensional}) for $N=5$ (thick red) and $N=10$ (dashed black). The upper-right, lower-left, and lower-right figures display the simulation results, where molecules exchange energies with random amounts, derived from analyzing ten thousand, one hundred thousand, and one million samples, respectively. Note that the results for $N=5$ (red) and $N=10$ (gray) overlap in the figures of the simulation results.}
\label{fig:fig3}
\end{figure}

We conducted computer simulations in which entities exchanged energies in random amounts. The results of these simulations are presented in Fig.~(\ref{fig:fig3}). The upper-left figure shows a reference calculated using Eq.~(\ref{eq:OneMoleculeVelocityAdimensional}) for $N=5$ and $10$, which is compared to the simulation results. The thick (red) line is for $N=5$, and the dashed (black) line is for $N=10$. The upper-right, lower-left, and lower-right figures show the simulation results obtained by analyzing ten thousand, one hundred thousand, and one million samples, respectively. Note that the results for $N=5$ (red) and $10$ (gray) overlap. The curves obtained using Eq.~(\ref{eq:OneMoleculeVelocityAdimensional}) are drawn in the lower-right figure. We observe that the theoretical result shows excellent agreement with the simulation result.

\begin{figure}
\centering
\includegraphics[scale=0.9]{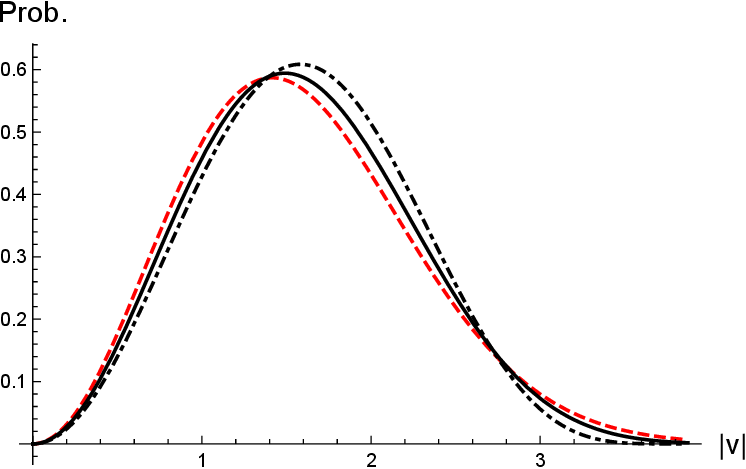}
\caption{Shown in this figure are the speed distributions ${\mathbf F}_{|\bar {\mathbf v}|}^3$, presented for comparison with the Maxwellian distribution. The dashed (red) line corresponds to the result of $N$ being infinity or the Maxwellian distribution. The solid (black) and the dot-dashed (black) lines, respectively, represent the results obtained for $N=5$ and $10$.}
\label{fig:fig4theory}
\end{figure}

From now, let us consider a system with $D$ degrees of freedom, such as the Maxwellian distribution with respect to the speed in three-dimensional space. This distribution is expressed in a neat form using dimensionless variables, as shown in Eq.~(\ref{eq:MaxwellianVelocityAdimensional}), by
\begin{equation}
\label{eqAdimensionalMB3Dspeed}
{\mathbf M_{|\bar {\mathbf v}|}}(|\bar {\mathbf v}|)  = \left(\frac{2}{\pi}\right)^{1/2} |\bar {\mathbf v}|^2  \exp \left(- \frac {|\bar {\mathbf v}|^2}{2} \right)
\end{equation}
where $\bar {\mathbf v} = (\bar v_x, \bar v_y, \bar v_z)$. Previously, we derived the energy distribution given by Eq.~(\ref{eq:OneMoleculeEnergy}), which approaches the Maxwellian energy distribution with one degree of freedom as $N$ increases. Extending this result to $D$ degrees of freedom is straightforward. The result is expressed by
\begin{equation}
\label{eqEnergyDistributionDdimensional}
{\mathbf F}_\epsilon (\epsilon) = \frac {\Gamma{\left( \frac{DN}{2}\right)} E^{-D/2} \epsilon^{(D-2)/2} }{\Gamma{\left(\frac D 2\right)} \Gamma {\left( \frac{D\left(N-1\right)}{2}\right)}} \left( 1-\frac {\epsilon}{E}\right)^{\frac {D(N-1)-2}{2}}
\end{equation}
where $0 \leq \epsilon \leq E$. One way to obtain Eq.~(\ref{eqEnergyDistributionDdimensional}) is to define 
$$\epsilon_i = q_i^2 = \sum_{p=1}^D q_{i,p}$$
where $q_{i,p}$ is defined by the polar coordinate system in the same manner to that of $q_i$ but with another set of variables $\varphi_p$ instead of $\phi_i$. Now if we substitute
$$E=\frac{D N k T}{2} $$
and
applying a change of variables in Eq.~(\ref {eqEnergyDistributionDdimensional}) as $$\epsilon = \frac {m  {|\mathbf v|}^2} {2} = \frac {kT}{2} {|\bar {\mathbf v}|}^2,$$
then we have
\begin{equation}
\label{eqSpeedDistributionDdimensionalGas}
{\mathbf F}_{|\bar {\mathbf v}|}^D  \left({|\bar {\mathbf v}|}\right) = 2 \frac {{\Gamma \left( \frac{DN}{2}\right)} {|\bar {\mathbf v}|}^{(D-1)} {\left({DN}\right)}^{-D/2} } {\Gamma{\left(\frac D 2\right)} \Gamma {\left( \frac{D\left(N-1\right)}{2}\right)}} \left( 1-\frac {|\bar {\mathbf v}|^2}{DN}\right)^{\frac {D(N-1)-2}{2}}
\end{equation}
and when $D=3$, we have compared ${\mathbf F}_{|\bar {\mathbf v}|}^3$ to the corresponding Maxwellian distribution Eq.~(\ref{eqAdimensionalMB3Dspeed}) in Fig.~(\ref{fig:fig4theory}). The dashed (red) line corresponds to the limiting case of $N$ being infinity, \textit{i.e.}, the Maxwellian distribution, while the solid (black) and the dot-dashed (black) lines represent the results of $N=5$ and $10$, respectively. Notably, the short tails observed in Fig.~(\ref{fig:fig4theory}) with respect to the Maxwellian distribution are consistent with experimental findings of molecular evaporation of acid dimers of \cite{Faubel} and desorption of hydrogen molecules from a copper crystal of \cite{jitschin&} where the figures of the references use flight time instead of velocity so that the positions of peaks are inversed, and normalize the heights of peaks.

\begin{figure}
\centering
\includegraphics[scale=0.9]{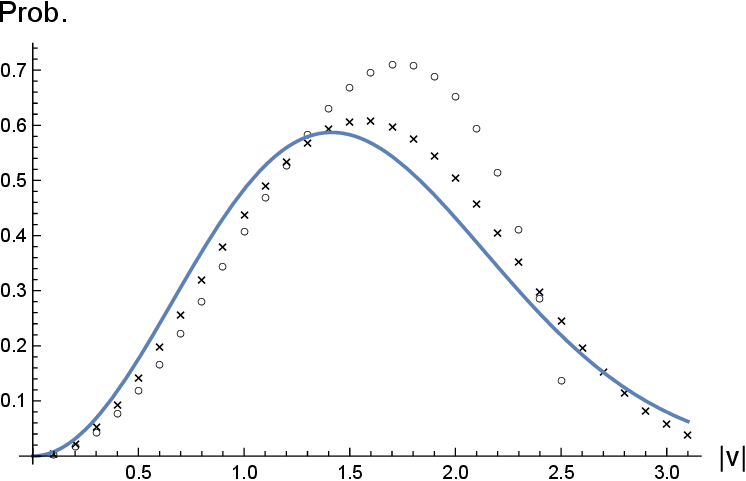}
\caption{The speed distributions obtained from $\tilde { \mathbf F}_{\bar{\mathbf v}} (\bar {\mathbf v})$ are shown in the figure. The (blue) line corresponds to the Maxwellian distribution obtained in the limit $N\rightarrow\infty$. The cross and the circular symbols correspond to the distributions obtained for $N=5$ and $10$, respectively.}
\label{fig:fig4}
\end{figure}

The non-Maxwellian distribution with respect to velocity can be obtained through a simple change of integration variable, as follows:
\begin{equation}
\label{eqVelocityDistributionDdimensionalGas}
{\mathbf F}_{\bar {\mathbf v}}^D  \left({\bar {\mathbf v}}\right) = \frac {{\Gamma \left( \frac{DN}{2}\right)}  {\left({\pi DN }\right)}^{-D/2} } { \Gamma {\left( \frac{D\left(N-1\right)}{2}\right)}} \left( 1-\frac {|\bar {\mathbf v}|^2}{DN}\right)^{\frac {D(N-1)-2}{2}}
\end{equation}
where the support is defined by $|\bar {\mathbf v}|^2 \leq DN$. One may obtain Eq.~(\ref{eqVelocityDistributionDdimensionalGas}) equation by using Havrda-Charv\'at-Tsallis entropy as shown by \cite{shim2020}. Note that the Maxwellian distribution in three-dimensional space $\mathbf M_{\bar{\mathbf v}}(\bar{\mathbf v})$ is obtained by just multiplying the Maxwellian distributions in one-dimensional space $\bar M(\bar v_x)$, $\bar M(\bar v_y)$, and $\bar M(\bar v_z)$ as
$$\mathbf M_{\bar{\mathbf v}}(\bar{\mathbf v})=\bar M(\bar v_x) \bar M(\bar v_y) \bar M(\bar v_z),$$
however, ${\mathbf F}_{\bar {\mathbf v}}^3  \left({\bar {\mathbf v}}\right)$ of Eq.~(\ref{eqVelocityDistributionDdimensionalGas}) with $D=3$ is different from the tensor product of the distributions of one degree of freedom as

\begin{align}
\label{eqAdimensional3Dspeed}
& \tilde { \mathbf F}_{\bar{\mathbf v}} (\bar {\mathbf v})  \nonumber \\
&=\left(\frac{\Gamma \left(\frac N 2 \right) }{\Gamma \left(\frac {N-1} 2 \right) \sqrt {\pi N}}\right)^3 \left(\left(1-\frac {\bar v_x^2}{N} \right)\left(1-\frac {\bar v_y^2}{N} \right)\left(1-\frac {\bar v_z^2}{N} \right)\right)^{\frac{N-3}{2}}.
\end{align}

The result of a numerical calculation of the speed distribution obtained by using Eq.~(\ref{eqAdimensional3Dspeed}) is shown in Fig.~(\ref{fig:fig4}). The (blue) line represents the Maxwellian distribution, where $N$ being infinity, while the cross and the circular symbols correspond to $N=5$ and $10$, respectively. The speed distribution was computed by generating quite uniformly distributed points on the surface of spheres with varying radii equivalent to speeds, and calculating probabilities by summing up the values of Eq.~(\ref{eqAdimensional3Dspeed}) for each set of points having the same radius. We emphasize that the directional independence of the velocity distribution is only valid in the Maxwellian distribution where $N$ is assumed to be infinity. Otherwise, the directional probability is dependent. Note that when $\bar {\mathbf v}=(0,0,0)$ and $N=6.02 \times 10^{23}$ which is the Avogadro's number, ${\mathbf F}_{\bar {\mathbf v}}^3  \left({\bar {\mathbf v}}\right) - \tilde { \mathbf F}_{\bar{\mathbf v}} (\bar {\mathbf v})=-1.05 \times 10^{-25}$ and ${\mathbf F}_{\bar {\mathbf v}}^3  \left({\bar {\mathbf v}}\right) - \mathbf M_{\bar{\mathbf v}}(\bar{\mathbf v}) = 3.59\times 10^{-8} $ which means the infinity assumption generates more error than the directional independence assumption even in a scale of the Avogadro's.  
 
\section{Conclusion}
\label{sec:conclusion}

In this study, we have derived distribution functions for a system with a finite number of entities. These distributions are non-Gaussian and non-chi-squared, but they approach the corresponding Gaussian distributions as the number of entities increases towards infinity. The parameter of the derived distribution is expressed by the number of entities. The corresponding entropy formula of the derived distribution is also derived with the parameter expressed by the number of entities.

The derived distributions may be employed in a diverse range of economic and financial frameworks where a limited number of agents engage in interactions, encompassing models of oligopoly, game theory, and financial markets. For a fundamental example, we consider a game where a group of entities, possessing identical competencies, participate in a repetitive exchange of a portion of their assets to arbitrarily selected counterparts. Within this setting, the asset allocation follows the non-chi-square distribution we derived in our manuscript. Accordingly, in biological domains, the derived non-Gaussian distributions show potential for modeling the behavior of organismic or cellular populations, especially in cases where the population is small and the entities interact.

\section*{Acknowledgement}
This work was partially supported by the KIST Institutional Program.

\bibliographystyle{chicagomine}
\bibliography{conservedSystemWithLimited}

\end{document}